\documentclass{article}

\PassOptionsToPackage{numbers, compress}{natbib}

\usepackage[final, nonatbib]{neurips_2021_ml4ps}
\usepackage[utf8]{inputenc} % allow utf-8 input
\usepackage[T1]{fontenc}    % use 8-bit T1 fonts
\usepackage{hyperref}       % hyperlinks
\usepackage{url}            % simple URL typesetting
\usepackage{booktabs}       % professional-quality tables
\usepackage{amsfonts}       % blackboard math symbols
\usepackage{nicefrac}       % compact symbols for 1/2, etc.
\usepackage{microtype}      % microtypography
\usepackage{xcolor}         % colors
\usepackage{graphicx}
\usepackage{multirow}
\usepackage{booktabs}

\title{E(2) Equivariant Self-Attention for Radio Astronomy}

\author{
  Micah Bowles \\
  Department of Physics \& Astronomy\\
  University of Manchester, UK\\
  \texttt{micah.bowles@postgrad.manchester.ac.uk} \\
  \And
  Matthew Bromley \\
  Department of Physics \& Astronomy\\
  University of Manchester, UK\\
 \texttt{matthew.bromley@manchester.ac.uk} \\
  \AND
  Max Allen \\
  Department of Physics \& Astronomy\\
  University of Manchester, UK\\
  \texttt{max.allen@manchester.ac.uk} \\
  \And
  Anna Scaife\thanks{The Alan Turing Institute, 96 Euston Rd, London, UK \texttt{a.scaife@turing.ac.uk}} \\
  Department of Physics \& Astronomy\\
  University of Manchester, UK\\
  \texttt{anna.scaife@manchester.ac.uk} \\
}

\begin{document}

\maketitle

\begin{abstract}
  In this work we introduce group-equivariant self-attention models to address the problem of explainable radio galaxy classification in astronomy. We evaluate various orders of both cyclic and dihedral equivariance, and show that including equivariance as a prior both reduces the number of epochs required to fit the data and results in improved performance. We highlight the benefits of equivariance when using self-attention as an explainable model and illustrate how equivariant models statistically attend the same features in their classifications as human astronomers.
\end{abstract}

\section{Introduction}\label{sec: Introduction}
A number of radio astronomy instruments have been and are being constructed which will produce daunting volumes of data, for which the astronomical community is not yet fully prepared \cite{SKADataProcessor2015}. One challenge the community faces is to classify the millions of individual objects that will be detected by these instruments \cite{VanHaarlem2013, Beardsley2019, Jarvis2016, Johnston2008}. Supervised deep learning is highly suited for many instances of this task as the data are highly dimensional with existing labels (e.\,g. \cite{fr1974}) for various types of objects assigned by astronomers throughout the history of radio observations.

In particular, Convolutional Neural Networks (CNNs) have been researched for a number of problems in radio astronomy, including the classification of radio galaxies, e.\,g.\cite{aniyan2017,hmtnet}.
To answer various scientific questions around the physical properties of radio galaxies and the conditions through which they form, they must first be catalogued. This requires astronomers to classify such objects into their sub-types, in particular FRI (edge darkened) and FRII (edge brightened) radio galaxies \cite{fr1974}. This task is made more challenging by noise, instrumental effects, and the substantial differences often seen between any two individual galaxies of the same class. Consequently, interpretation of individual model predictions is of particular interest to the astronomy community. With this in mind, \cite{bowles2020} showed that a feature-wise self-attention model could not only be used to make \textit{explainable} predictions for an audience of astronomers by producing attention (saliency) maps, but that it also performed similarly to other state-of-the-art radio galaxy classifiers whilst using 50\% fewer parameters than the next smallest classic CNN application in the field. 

In this work, following \cite{bowles2020, e2cnn}, we introduce \emph{group-equivariant} self-attention gates for classification of FRI and FRII radio galaxies. Incorporating prior constraints on rotational and reflectional equivariance is a topic of growing interest in astronomy (e.\,g. \cite{dieleman2015,ntwaetsile,e2cnn}) where the orientation and chirality of astronomical systems is considered to be largely unbiased. Here we use the \verb+e2cnn+ package introduced in \cite{weilercesa2019} to impose group equivariance priors on our self-attention models. We impose cyclical (C; rotation) and dihedral (D; rotation and reflection) two-dimensional Euclidean, E(2), equivariance. Each of the equivariance priors in our model are limited to an integer number of discrete rotations, either 4, 8 or 16 in this work, where C4 means the operations are equivariant to rotations of $0$, $\pm\frac{\pi}{2}$, and $\pi$ radians.

\section{Models}
\label{sec:Models}

We investigate two architectures in this work. First, to allow for direct comparisons with \cite{e2cnn}, we train a LeNet \cite{lenet} style architecture without equivariance. Second, the architecture of primary interest in this work is the self-attention model used in \cite{bowles2020}, but with a kernel size of 5 to allow for the equivariance constraints to be modeled correctly. The equivariance of a given model is indicated as cyclic (C) or dihedral (D), with an order of rotational symmetry, e.g. D8.

Our implementation of additive self-attention follows \cite{schlemper2018attention, vaswani2017attention} and is a feature wise transformer using feature planes of the globally encoded output, $g\in \mathbb{R}^{C_g \times H_g \times W_g}$, and a given input, $x\in\mathbb{R}^{C_x \times H_x \times W_x}$. This implementation produces an attention map, $\alpha\in[0,1]^{1 \times H_g \times W_g}$, and an attended output, $y\in \mathbb{R}^{C_x \times H_x \times W_x}$, and can be summarised as
\begin{equation}\label{eqn: attention map}
    \alpha = \sigma\left[W\circ relu\left(W_x(x) + f\circ W_g(g)\right)\right],
\end{equation}
\begin{equation}\label{eqn: attended output}
y_{ijk} = x_{ijk} \cdot \alpha_{1jk},
\end{equation}
where three 1$\times$1-convolution operations are defined through their dimensions as
$W: \mathbb{R}^{C_k \times H_x \times W_x} \rightarrow \mathbb{R}^{1 \times H_x \times W_x}$, 
$W_g: \mathbb{R}^{C_g \times H_g \times W_g} \rightarrow \mathbb{R}^{C_k \times H_g \times W_g}$, and
$W_x: \mathbb{R}^{C_g \times H_g \times W_g} \rightarrow \mathbb{R}^{C_k \times H_x \times W_x}$. 
Additionally, $f:\mathbb{R}^{C_g \times H_g \times W_g}$ is a (bilinear) upsampling function, $relu(x) = max\{0,x\}$ is the element-wise rectified linear unit,
and we define a normalisation
$\sigma: \mathbb{R}^{1 \times H_x \times W_x} \rightarrow \mathbb{R}^{1 \times H_x \times W_x}$. To maintain consistency with \cite{bowles2020}, the normalisation is selected to be range normalisation: $\sigma(x) = [x - \min(x)]/[\max(x)-\min(x)]$.

Common CNN classifiers often make use of a number of wide fully connected layers at the end of their convolutional framework (e.\,g. \cite{simonyan2014very,lenet}), which are able to produce highly non-linear outputs. For self-attention as an explainable deep learning (XAI) tool, the classifications made by the highly non-linear fully connected layers would not be well-represented in the attention map. Therefore, to maximise the value of the attention maps as an explanatory tool for classification, we choose to minimise the number of parameters after the attention mechanisms. We use the fine tuned aggregation method detailed in \cite{bowles2020}.

\textbf{Training.} Our models are trained on a binary version of the publicly available MiraBest\footnote{Data are publicly available under a Creative Commons 4.0 license at \url{https://doi.org/10.5281/zenodo.4288836}} data set \cite{mirabest}, which contains 1225 images of FRI and FRII galaxies (856 train, 213 validation, and 153 test samples). Each model is trained using the Adam optimiser \cite{adam} with a learning rate of $10^{-5}$, weight decay and mini-batching with a batch size of 16. We augment the data with rotation, reflection and minor scaling (in the range of 0.9 and 1.1 times the image size). The model weights are updated for each of eight random data set augmentations in a single epoch. We evaluate the model at the epoch with the lowest validation loss.

Models\footnote{The full code is publicly available under GNU General Public License v3.0 at \url{https://github.com/mb010/EquivariantSelfAttention}} were trained using an Nvidia A100 GPU. The selected architecture strongly influences how long the training takes: 4\,h LeNet, 14\,h Attention, 19\,h C4-, 20\,h C8-, 33\,h C16-, 16\,h D4-, 37\,h D8-, 110\,h D16-Attention. The models we present in this work are the models within the training that present the lowest validation loss. Therefore, training can be shortened significantly by implementing early stopping criteria, but in our case we were interested in how the training behaved even after validation loss began to increase again (see Figure~\ref{fig:Equivariant Training}).

\section{Results}
\label{sec:results}

The performances of various models trained with different levels of equivariance are shown in Table~\ref{tab: evaluation of models}. Here we use the combined Mirabest dataset including data samples qualified as both Confident and Uncertain and we see a significant accuracy ceiling imposed on the models through MiraBest's label noise. This becomes clear when considering the increased performance of models trained and evaluated only on the Confident MiraBest labels, where for example in \cite{e2cnn} the non-equivariant LeNet architecture (equivalent to that trained in this work) achieved a 94\% test accuracy.
\begin{table}[ht]
\centering
\caption{Performance metrics for all models considered in this work.}
\label{tab: evaluation of models}
\begin{tabular}{@{}rcrrrrrrr@{}}
\toprule
\multicolumn{1}{c}{\multirow{2}{*}{\textbf{Model}}} &
  \multicolumn{1}{c}{\multirow{2}{*}{\textbf{Accuracy (\%)}}} &
  \multicolumn{1}{c}{\multirow{2}{*}{\textbf{AUC}}} &
  \multicolumn{2}{c}{\textbf{F1 Score}} &
  \multicolumn{2}{c}{\textbf{Precision}} &
  \multicolumn{2}{c}{\textbf{Recall}} \\ \cmidrule(l){4-9} 
\multicolumn{1}{c}{} &
  \multicolumn{1}{c}{} &
  \multicolumn{1}{c}{} &
  \multicolumn{1}{c}{\textbf{FRI}} &
  \multicolumn{1}{c}{\textbf{FRII}} &
  \multicolumn{1}{c}{\textbf{FRI}} &
  \multicolumn{1}{c}{\textbf{FRII}} &
  \multicolumn{1}{c}{\textbf{FRI}} &
  \multicolumn{1}{c}{\textbf{FRII}} \\ \midrule
LeNet         & 82.9          & 0.905          & 0.821          & 0.836          & 0.832          & 0.827          & 0.811          & 0.846          \\
Attention     & 83.8          & 0.909          & 0.819          & 0.854          & 0.894          & 0.800          & 0.755          & 0.916          \\
C4-Attention  & 83.0          & 0.921 & \textbf{0.833} & 0.826          & 0.792          & \textbf{0.873} & \textbf{0.879} & 0.784          \\
C8-Attention  & 82.6          & 0.871          & 0.823          & 0.829          & 0.811          & 0.841          & 0.834          & 0.818          \\
C16-Attention & 82.8          & 0.880          & 0.812          & 0.842          & 0.862          & 0.803          & 0.768          & 0.885          \\
D4-Attention  & 84.1          & 0.889          & 0.828          & 0.853          & 0.870          & 0.819          & 0.790          & 0.889          \\
D8-Attention  & \textbf{84.8} & 0.911          & 0.829          & \textbf{0.864} & \textbf{0.911} & 0.806          & 0.761          & \textbf{0.931} \\
D16-Attention & 81.8          & \textbf{0.923} & 0.813          & 0.823          & 0.808          & 0.827          & 0.818          & 0.818          \\ \bottomrule
\end{tabular}
\end{table}

Collectively the equivariant models marginally outperform the non-equivariant models. We make special note of the D16-Attention model which performs worse than the LeNet model in all classification metrics with the exception of the AUC, this is likely due to the severity of the D16 prior imposed on the 5x5-kernel over smoothing the features of the source which may otherwise have been used for classification \cite{e2cnn}. Overall, the uncertainties associated with label noise and the test data size do not allow for a significant separation of the equivariant models from the non-equivariant models. 

\begin{figure}[h]
    \centering
    \includegraphics[width=\linewidth]{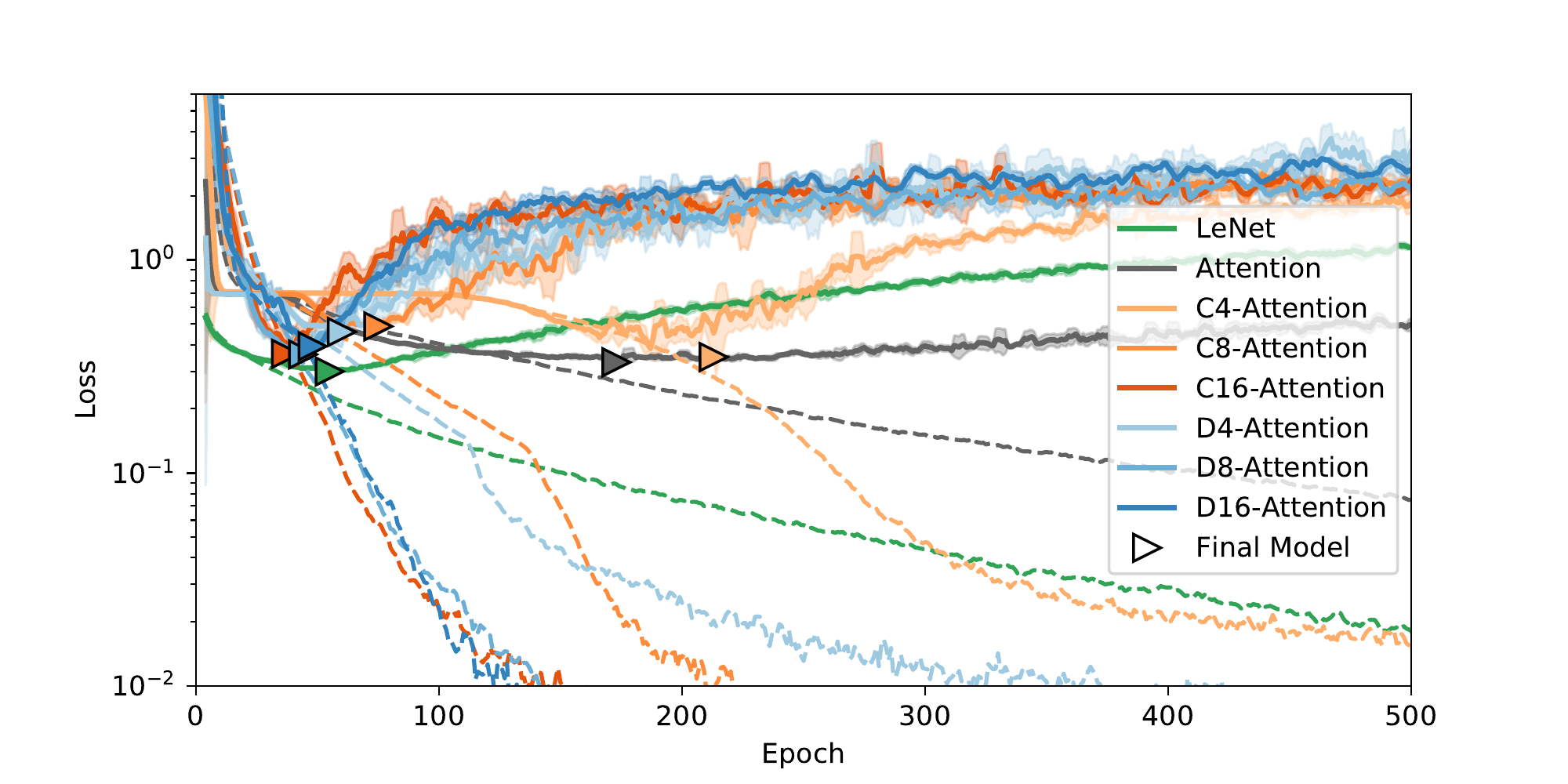}
    \caption{Validation (solid lines) and training (dashed lines) loss curves for all self-attention models listed in Table~\ref{tab: evaluation of models}. The respective 'Final Model' arrow heads highlight the respective models with the lowest validation loss used as the final models in this work. Curves are smoothed using a rolling window of 5 epochs, whose standard deviation is presented as the shaded region.}
    \label{fig:Equivariant Training}
\end{figure}

To understand how the equivariant models train, we consider the training loss curves for all models, see Figure~\ref{fig:Equivariant Training}. Note that the equivariant models converge towards a lower minimum in fewer epochs within the training loss landscape. We suggest that this is because with a correctly selected prior the models have a lower dimensional data space to map.

\begin{figure}
    \centering
    \includegraphics[width=\linewidth]{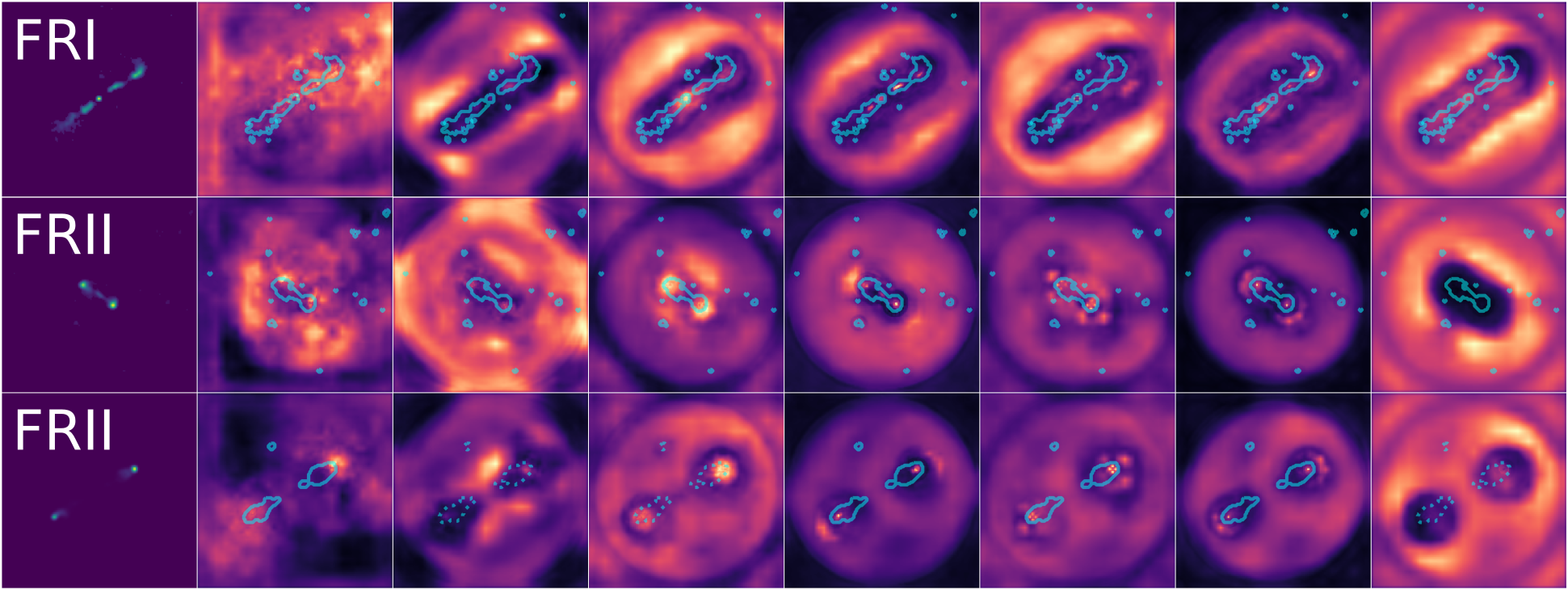}
    \caption{From left to right: The input source, followed by the respective attention maps of the presented models in the following order: non-equivariant, C4, C8, C16, D4, D8, and D16 models respectively. Contours show the zero level in the input to aid in identifying relevant structures, and are dashed if the models resulting classification was incorrect.}
    \label{fig:Source Amaps}
\end{figure}

As an explainable AI tool, we highlight the value of equivariance for providing additional stability and relatability in the resulting attention maps. Figure~\ref{fig:Source Amaps} highlights that attention maps which are equivariant are easier to interpret due to their smooth features and the equivariant treatment of the features within the image. It can be seen that the equivariant model extracts the brightest points of the respective lobes and jets of the radio source more clearly in comparison to the non-equivariant model. However, we also note the presence of low-level residual structure in the equivariant attention maps, with symmetry corresponding to the order of the kernel. We suggest that this may be due to the small support of the kernels leading to discretization artefacts. 

\begin{figure}
    \centering
    \includegraphics[width=0.75\linewidth]{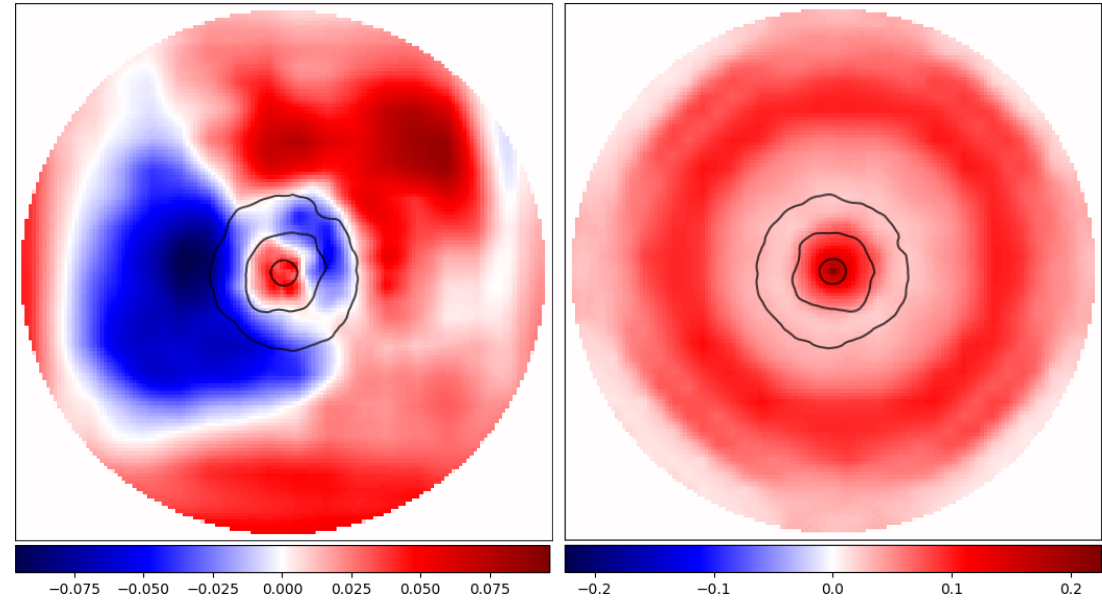}
    \caption{Class differenced mean pixel attentions for the non-equivariant model (left) and the D8 model (right). Contours highlight regions in which the one class begins to dominate across the augmented test set (i.e. contours highlight values at 0 for the differenced mean pixel augmented test data values).}
    \label{fig:Amap Distributions}
\end{figure}

Shown in Figure~\ref{fig:Amap Distributions} is the mean differenced attention between classes across the test sample, to highlight statistical spatial differences in the attention maps. The average radial amplitude differences in the MiraBest test data are marked as black contours on the image, and we note that although both models highlight the central region, the non-equivariant model has a statistically significant asymmetry in its mean attention maps, and therefore in its learned convolutional operations, which does not reflect how an expert human classifier would approach this problem.

\section{Conclusion}\label{sec:Conclussion}

In this work we have explored the use of group-equivariant self-attention as a tool for radio astronomy classification, where small labelled data sets may benefit from additional a priori constraints. We find that equivariant models train in fewer epochs and perform slightly better than their non-equivariant equivalents. Moreover, we find that group-equivariant self-attention models produce attention maps that seem to align better than their non-equivariant counterparts with the structures that would be attended by an expert human classifier. However, we note that whilst these results are encouraging there are a number of areas where more detailed investigations are still required. In particular a more extensive and quantitative approach to evaluating generalisation for larger test data sets is required, as well as an investigation into the residual structures observed in attention maps from equivariant convolutions thought to be caused by finite kernel support.  

% Remove ack environment after review process for final version.
\begin{ack}
The authors gratefully acknowledge support from the UK Alan Turing Institute under grant reference EP/V030302/1. MB gratefully acknowledges support from the UK Science \& Technology Facilities Council (STFC).
\end{ack}

% everything above here ^^^^ has to fit on 5 pages

\clearpage

\bibliography{references.bib}

\begin{thebibliography}{10}

\bibitem{aniyan2017}
A.~K. Aniyan and K.~Thorat.
\newblock {Classifying Radio Galaxies with the Convolutional Neural Network}.
\newblock {\em The Astrophysical Journal Supplement Series}, 2017.

\bibitem{Beardsley2019}
A.~P. Beardsley, M.~Johnston-Hollitt, C.~M. Trott, J.~C. Pober, J.~Morgan,
  D.~Oberoi, D.~L. Kaplan, C.~R. Lynch, G.~E. Anderson, P.~I. McCauley,
  S.~Croft, C.~W. James, O.~I. Wong, C.~D. Tremblay, R.~P. Norris, I.~H.
  Cairns, C.~J. Lonsdale, P.~J. Hancock, B.~M. Gaensler, N.~D. Bhat, W.~Li,
  N.~Hurley-Walker, J.~R. Callingham, N.~Seymour, S.~Yoshiura, R.~C. Joseph,
  K.~Takahashi, M.~Sokolowski, J.~C. Miller-Jones, J.~V. Chauhan,
  I.~Boji{\v{c}}i{\'{c}}, M.~D. Filipovi{\'{c}}, D.~Leahy, H.~Su, W.~W. Tian,
  S.~J. McSweeney, B.~W. Meyers, S.~Kitaeff, T.~Vernstrom, G.~G{\"{u}}rkan,
  G.~Heald, M.~Xue, C.~J. Riseley, S.~W. Duchesne, J.~D. Bowman, D.~C. Jacobs,
  B.~Crosse, D.~Emrich, T.~M. Franzen, L.~Horsley, D.~Kenney, M.~F. Morales,
  D.~Pallot, K.~Steele, S.~J. Tingay, M.~Walker, R.~B. Wayth, A.~Williams, and
  C.~Wu.
\newblock {Science with the Murchison Widefield Array: Phase i results and
  Phase II opportunities}.
\newblock {\em Publications of the Astronomical Society of Australia}, 2019.

\bibitem{bowles2020}
M.~{Bowles}, A.~M.~M. {Scaife}, F.~{Porter}, H.~{Tang}, and D.~J. {Bastien}.
\newblock {Attention-gating for improved radio galaxy classification}.
\newblock {\em Monthly Notices of the Royal Astronomical Society},
  501(3):4579--4595, Mar. 2021.

\bibitem{SKADataProcessor2015}
P.~Broekema, R.~Van~Nieuwpoort, and H.~Bal.
\newblock The square kilometre array science data processor. preliminary
  compute platform design.
\newblock {\em Journal of Instrumentation}, 10:C07004--C07004, 07 2015.

\bibitem{dieleman2015}
S.~Dieleman, K.~W. Willett, and J.~Dambre.
\newblock {Rotation-invariant convolutional neural networks for galaxy
  morphology prediction}.
\newblock {\em Monthly Notices of the Royal Astronomical Society},
  450(2):1441--1459, 04 2015.

\bibitem{fr1974}
B.~L. {Fanaroff} and J.~M. {Riley}.
\newblock {The morphology of extragalactic radio sources of high and low
  luminosity}.
\newblock {\em Monthly Notices of the Royal Astronomical Society},
  167:31P--36P, May 1974.

\bibitem{Jarvis2016}
M.~J. Jarvis, A.~R. Taylor, I.~Agudo, J.~R. Allison, R.~P. Deane, B.~Frank,
  N.~Gupta, I.~Heywood, N.~Maddox, K.~McAlpine, M.~G. Santos, A.~M. Scaife,
  M.~Vaccari, J.~T. Zwart, E.~Adams, D.~J. Bacon, A.~J. Baker, B.~A. Bassett,
  P.~N. Best, R.~Beswick, S.~Blyth, M.~L. Brown, M.~Br{\"{u}}ggen, M.~Cluver,
  S.~Colafranceso, G.~Cotter, C.~Cress, R.~Dav{\'{e}}, C.~Ferrari, M.~J.
  Hardcastle, C.~Hale, I.~Harrison, P.~W. Hatfield, H.~R. Kl{\"{o}}ckner,
  S.~Kolwa, E.~Malefahlo, T.~Marubini, T.~Mauch, K.~Moodley, R.~Morganti,
  R.~Norris, J.~A. Peters, I.~Prandoni, M.~Prescott, S.~Oliver, N.~Oozeer,
  H.~J. R{\"{o}}ttgering, N.~Seymour, C.~Simpson, O.~Smirnov, D.~J. Smith,
  K.~Spekkens, J.~Stil, C.~Tasse, K.~van~der Heyden, I.~H. Whittam, and W.~L.
  WIlliams.
\newblock {The MeerKAT international GHz tiered extragalactic exploration
  (MIGHTEE) survey}.
\newblock In {\em Proceedings of Science}, 2016.

\bibitem{Johnston2008}
S.~Johnston, R.~Taylor, M.~Bailes, N.~Bartel, C.~Baugh, M.~Bietenholz,
  C.~Blake, R.~Braun, J.~Brown, S.~Chatterjee, J.~Darling, A.~Deller,
  R.~Dodson, P.~Edwards, R.~Ekers, S.~Ellingsen, I.~Feain, B.~Gaensler,
  M.~Haverkorn, G.~Hobbs, A.~Hopkins, C.~Jackson, C.~James, G.~Joncas,
  V.~Kaspi, V.~Kilborn, B.~Koribalski, R.~Kothes, T.~Landecker, A.~Lenc,
  J.~Lovell, J.~P. MacQuart, R.~Manchester, D.~Matthews, N.~McClure-Griffiths,
  R.~Norris, U.~L. Pen, C.~Phillips, C.~Power, R.~Protheroe, E.~Sadler,
  B.~Schmidt, I.~Stairs, L.~Staveley-Smith, J.~Stil, S.~Tingay, A.~Tzioumis,
  M.~Walker, J.~Wall, and M.~Wolleben.
\newblock {Science with ASKAP : The Australian square-kilometre-array
  pathfinder}.
\newblock {\em Experimental Astronomy}, 2008.

\bibitem{adam}
D.~P. {Kingma} and J.~{Ba}.
\newblock {Adam: A Method for Stochastic Optimization}.
\newblock {\em arXiv e-prints}, page arXiv:1412.6980, Dec. 2014.

\bibitem{lenet}
Y.~LeCun, L.~Bottou, Y.~Bengio, and P.~Haffner.
\newblock Gradient-based learning applied to document recognition.
\newblock {\em Proceedings of the IEEE}, 86(11):2278--2324, November 1998.

\bibitem{ntwaetsile}
K.~{Ntwaetsile} and J.~E. {Geach}.
\newblock {Rapid sorting of radio galaxy morphology using Haralick features}.
\newblock {\em Monthly Notices of the Royal Astronomy Scoiety},
  502(3):3417--3425, Apr. 2021.

\bibitem{mirabest}
F.~A.~M. Porter.
\newblock Mirabest batched dataset 10.5281/zenodo.4288837, Nov. 2020.

\bibitem{e2cnn}
A.~M.~M. {Scaife} and F.~{Porter}.
\newblock {Fanaroff-Riley classification of radio galaxies using
  group-equivariant convolutional neural networks}.
\newblock {\em Monthly Notices of the Royal Astronomical Society},
  503(2):2369--2379, May 2021.

\bibitem{schlemper2018attention}
J.~Schlemper, O.~Oktay, L.~Chen, J.~Matthew, C.~Knight, B.~Kainz, B.~Glocker,
  and D.~Rueckert.
\newblock Attention-gated networks for improving ultrasound scan plane
  detection.
\newblock {\em arXiv preprint arXiv:1804.05338}, 2018.

\bibitem{simonyan2014very}
K.~Simonyan and A.~Zisserman.
\newblock Very deep convolutional networks for large-scale image recognition.
\newblock {\em arXiv preprint arXiv:1409.1556}, 2014.

\bibitem{hmtnet}
H.~{Tang}, A.~M.~M. {Scaife}, and J.~P. {Leahy}.
\newblock {Transfer learning for radio galaxy classification}.
\newblock {\em Monthly Notices of the Royal Astronomical Society},
  488(3):3358--3375, Sept. 2019.

\bibitem{VanHaarlem2013}
M.~P. Van~Haarlem, M.~W. Wise, A.~W. Gunst, G.~Heald, J.~P. McKean, J.~W.
  Hessels, A.~G. De~Bruyn, R.~Nijboer, J.~Swinbank, R.~Fallows, M.~Brentjens,
  A.~Nelles, R.~Beck, H.~Falcke, R.~Fender, J.~H{\"{o}}randel, L.~V. Koopmans,
  G.~Mann, G.~Miley, H.~R{\"{o}}ttgering, B.~W. Stappers, R.~A. Wijers,
  S.~Zaroubi, M.~Van Den~Akker, A.~Alexov, J.~Anderson, K.~Anderson,
  A.~Van~Ardenne, M.~Arts, A.~Asgekar, I.~M. Avruch, F.~Batejat,
  L.~B{\"{a}}hren, M.~E. Bell, M.~R. Bell, I.~Van~Bemmel, P.~Bennema, M.~J.
  Bentum, G.~Bernardi, P.~Best, L.~B{\^{i}}rzan, A.~Bonafede, A.~J. Boonstra,
  R.~Braun, J.~Bregman, F.~Breitling, R.~H. Van De~Brink, J.~Broderick, P.~C.
  Broekema, W.~N. Brouw, M.~Br{\"{u}}ggen, H.~R. Butcher, W.~Van~Cappellen,
  B.~Ciardi, T.~Coenen, J.~Conway, A.~Coolen, A.~Corstanje, S.~Damstra,
  O.~Davies, A.~T. Deller, R.~J. Dettmar, G.~Van~Diepen, K.~Dijkstra,
  P.~Donker, A.~Doorduin, J.~Dromer, M.~Drost, A.~Van~Duin, J.~Eisl{\"{o}}ffel,
  J.~Van~Enst, C.~Ferrari, W.~Frieswijk, H.~Gankema, M.~A. Garrett,
  F.~De~Gasperin, M.~Gerbers, E.~De~Geus, J.~M. Grie{\ss}meier, T.~Grit,
  P.~Gruppen, J.~P. Hamaker, T.~Hassall, M.~Hoeft, H.~A. Holties, A.~Horneffer,
  A.~Van Der~Horst, A.~Van~Houwelingen, A.~Huijgen, M.~Iacobelli, H.~Intema,
  N.~Jackson, V.~Jelic, A.~De~Jong, E.~Juette, D.~Kant, A.~Karastergiou,
  A.~Koers, H.~Kollen, V.~I. Kondratiev, E.~Kooistra, Y.~Koopman, A.~Koster,
  M.~Kuniyoshi, M.~Kramer, G.~Kuper, P.~Lambropoulos, C.~Law, J.~Van~Leeuwen,
  J.~Lemaitre, M.~Loose, P.~Maat, G.~Macario, S.~Markoff, J.~Masters, R.~A.
  McFadden, D.~McKay-Bukowski, H.~Meijering, H.~Meulman, M.~Mevius,
  E.~Middelberg, R.~Millenaar, J.~C. Miller-Jones, R.~N. Mohan, J.~D. Mol,
  J.~Morawietz, R.~Morganti, D.~D. Mulcahy, E.~Mulder, H.~Munk, L.~Nieuwenhuis,
  R.~Van~Nieuwpoort, J.~E. Noordam, M.~Norden, A.~Noutsos, A.~R. Offringa,
  H.~Olofsson, A.~Omar, E.~Orr{\'{u}}, R.~Overeem, H.~Paas, M.~Pandey-Pommier,
  V.~N. Pandey, R.~Pizzo, A.~Polatidis, D.~Rafferty, S.~Rawlings, W.~Reich,
  J.~P. De~Reijer, J.~Reitsma, G.~A. Renting, P.~Riemers, E.~Rol, J.~W. Romein,
  J.~Roosjen, M.~Ruiter, A.~Scaife, K.~Van Der~Schaaf, B.~Scheers,
  P.~Schellart, A.~Schoenmakers, G.~Schoonderbeek, M.~Serylak, A.~Shulevski,
  J.~Sluman, O.~Smirnov, C.~Sobey, H.~Spreeuw, M.~Steinmetz, C.~G. Sterks,
  H.~J. Stiepel, K.~Stuurwold, M.~Tagger, Y.~Tang, C.~Tasse, I.~Thomas,
  S.~Thoudam, M.~C. Toribio, B.~Van Der~Tol, O.~Usov, M.~Van~Veelen, A.~J. Van
  Der~Veen, S.~Ter~Veen, J.~P. Verbiest, R.~Vermeulen, N.~Vermaas, C.~Vocks,
  C.~Vogt, M.~De~Vos, E.~Van Der~Wal, R.~Van~Weeren, H.~Weggemans,
  P.~Weltevrede, S.~White, S.~J. Wijnholds, T.~Wilhelmsson, O.~Wucknitz,
  S.~Yatawatta, P.~Zarka, A.~Zensus, and J.~Van~Zwieten.
\newblock {LOFAR: The low-frequency array}, 2013.

\bibitem{vaswani2017attention}
A.~Vaswani, N.~Shazeer, N.~Parmar, J.~Uszkoreit, L.~Jones, A.~N. Gomez,
  {\L}.~Kaiser, and I.~Polosukhin.
\newblock Attention is all you need.
\newblock In {\em Advances in neural information processing systems}, pages
  5998--6008, 2017.

\bibitem{weilercesa2019}
M.~Weiler and G.~Cesa.
\newblock General e(2)-equivariant steerable cnns.
\newblock In H.~Wallach, H.~Larochelle, A.~Beygelzimer, F.~d\textquotesingle
  Alch\'{e}-Buc, E.~Fox, and R.~Garnett, editors, {\em Advances in Neural
  Information Processing Systems}, volume~32. Curran Associates, Inc., 2019.

\end{thebibliography}
\bibliographystyle{abbrv}

\end{document}